\documentstyle[prb,aps,twocolumn]{revtex}
\input epsf
\begin{document}
\twocolumn[\hsize\textwidth\columnwidth\hsize\csname
@twocolumnfalse\endcsname

\draft

\title{Quantum Monte Carlo calculations of the one-body
density matrix and excitation energies of silicon}

\author{P.~R.~C.~Kent, Randolph~Q.~Hood, M.~D.~Towler, R.~J.~Needs,
and G.~Rajagopal}

\address{Cavendish Laboratory, Madingley Road, Cambridge CB3 0HE,
United Kingdom.}

\date{Received 8 October 1997}

\maketitle
\begin{abstract}
\begin{quote}
  \parbox{16cm}{\small Quantum Monte Carlo (QMC) techniques are used
    to calculate the one-body density matrix and excitation energies
    for the valence electrons of bulk silicon.  The one-body density
    matrix and energies are obtained from a Slater-Jastrow wave
    function with a determinant of local density approximation (LDA)
    orbitals. The QMC density matrix evaluated in a basis of LDA
    orbitals is strongly diagonally dominant.  The natural orbitals
    obtained by diagonalizing the QMC density matrix resemble the LDA
    orbitals very closely.  Replacing the determinant of LDA orbitals
    in the wave function by a determinant of natural orbitals makes no
    significant difference to the quality of the wave function's nodal
    surface, leaving the diffusion Monte Carlo energy unchanged.  The
    Extended Koopmans' Theorem for correlated wave functions is used
    to calculate excitation energies for silicon, which are in
    reasonable agreement with the available experimental data.  A
    diagonal approximation to the theorem, evaluated in the basis of
    LDA orbitals, works quite well for both the quasihole and
    quasielectron states.  We have found that this approximation has
    an advantageous scaling with system size, allowing more efficient
    studies of larger systems. }

\end{quote}
\end{abstract}


]


\section{Introduction}

The two most common, practical quantum Monte Carlo (QMC) methods for
realistic systems are the variational quantum Monte Carlo
(VMC)\cite{vmc,hammond} and diffusion quantum Monte Carlo
(DMC)\cite{hammond,dmc} methods. In VMC, expectation values are
computed with an approximate many-body trial wave function.  In DMC,
imaginary time evolution of the many-body Schr\"{o}dinger equation in
principle gives exact results, although in practice one needs to make
the ``fixed-node approximation'' to account for the antisymmetry of
the many-electron wave function. In the fixed node approximation, the
nodes of the propagated wave function are restricted to those of the
trial wave function. The accuracy of this approximation is central to
DMC simulations of many-electron systems. One of the aims of our work
is to investigate the effectiveness of this approximation for extended
systems, with the long term goal of obtaining better trial wave
functions.

In this paper we calculate the one-body density matrix for the valence
electrons of silicon within the VMC framework, and obtain the natural
orbitals which diagonalize the density matrix.  These calculations
require the whole of the density matrix throughout all of the
six-dimensional space ${\bf r} \times {\bf r'}$, not just at a few
points in space as has been obtained before.  To our knowledge this is
the first time that the one-body density matrix and natural orbitals
have been obtained for an extended, inhomogeneous, interacting
electron system.  Recent evidence has suggested\cite{jcg:sicluster}
that a determinant of natural orbitals may give a better nodal surface
than a determinant of Hartree-Fock (HF) orbitals. Our results show
that a determinant of natural orbitals has a similar quality nodal
surface to a determinant of LDA orbitals for bulk silicon. In a
separate calculation we find that a determinant of LDA orbitals has a
slightly better nodal surface than a determinant of HF orbitals.

There is considerable interest in calculating excitation energies
using QMC techniques.  Excitation energies may be obtained by
analyzing DMC decay curves,\cite{mc:releasenode,dmc:decay} but this
method has not proven very useful due to the large statistical noise.
Furthermore, as the quality of the ground state trial wave function
improves, less information about excited states is obtained. A
combination of ground and excited state wave functions must then be
used to obtain upper bounds for the excitation energies.  Direct
methods for calculating excitation energies have met with more
success. Mit\'{a}\v{s} and Martin have calculated an excitation energy
in a molecular nitrogen solid by performing DMC calculations for the
ground and excited states.\cite{lm:nitrogen} Mit\'{a}\v{s} has also
reported similar calculations for two excitation energies in
diamond.\cite{lm:compcarb} Recently\cite{ourbandgapletter} we used the
same method to calculate 27 excitation energies in silicon, obtaining
very good agreement with experiment for the low lying excitation
energies, while the energies of the higher lying excitations were
somewhat too large.  In this paper we calculate excitation energies
using a different approach.  Here we use the ``Extended Koopmans'
Theorem'' (EKT),\cite{owd:ektpotl,mmm:ekt} which derives from quantum
chemistry, and involves the one-body density matrix.  We have applied
this theorem within VMC to calculate the excitation energies of
silicon at four inequivalent {\bf k}-points within the Brillouin zone.
The energies are in good agreement with the available experimental
data with a level of agreement similar to direct excitation
calculations.  We also test the diagonal approximation to the EKT
evaluated using the LDA orbitals, which was used previously to
estimate quasihole energies in silicon\cite{sf:pairprl} and
NiO.\cite{st:nio2} We find that the approximation performs well in
silicon and that it has an advantageous scaling with system size.
This allows more efficient studies of excitations in large systems
than are possible with existing direct techniques.

The layout of this paper is as follows.  In section \ref{sec:qmc} we
briefly describe the QMC techniques used in our calculations,
including the Hamiltonian, the trial wave function and the relevance
of natural orbitals to QMC calculations.  In section \ref{sec:dm} we
present and discuss our results for the one-body density matrix and
the natural orbitals of silicon.  In section \ref{sec:ee} we describe
the Extended Koopmans' Theorem and its application to the band
structure of silicon.

\section{QMC Simulations of Silicon}
\label{sec:qmc}
In this section we briefly describe our QMC calculations.  For a more
detailed discussion of the methods we refer the reader to the
literature.\cite{hammond,dmc,sf:prb,ouroptim,ourfinitesize}

\subsection{The Hamiltonian}

For this study we used an fcc simulation cell, with periodic boundary
conditions, containing 54 Si$^{4+}$ ions and 216 electrons. The
Hamiltonian for our system, within the Born-Oppenheimer approximation,
is

\begin{eqnarray}
\label{eqn:hamil}
 \hat{H} & = & \sum_{i}-\frac{1}{2}\nabla^2_i +
 \sum_i\sum_{\alpha}v_{\alpha}({\bf r}_i,{\bf d}_\alpha) \nonumber \\
 & + & \frac{1}{2}\sum_i\sum_{j\neq i}v({\bf r}_i,{\bf r}_j) +
 \frac{1}{2}\sum_{\alpha}\sum_{\beta\neq \alpha}v_{\alpha\beta}({\bf
 d}_{\alpha},{\bf d}_{\beta})\;\;.
\end{eqnarray}

\noindent The positions of the $N$ electrons in the supercell are
denoted by ${\bf r}_i$ and the ion locations are denoted by ${\bf
d}_\alpha$.  The electron-ion potential, $v_{\alpha}$, is modeled by a
norm-conserving non-local pseudopotential\cite{nt:pseudo} obtained
from atomic calculations performed within the local density
approximation (LDA) to density functional theory. The standard method
for including the inter-particle Coulomb interactions in periodic
systems is to use the Ewald interaction potential.  We have found that
this interaction gives rise to significant finite size errors,
especially for small simulation cells.  Recently we introduced a new
formulation of the electron-electron interaction for simulations using
periodic boundary conditions which eliminates this
problem\cite{ourfinitesize} (hereafter referred to as the ``cutoff
interaction''). This interaction satisfies the conditions that (i) it
gives the correct Hartree energy and (ii) it has the proper $1/r$ form
for the interaction of an electron with its exchange-correlation
hole. (The Ewald interaction violates condition (ii).) Here we present
results for excitation energies calculated with both the Ewald and
cutoff interactions, using a wave function which was optimized using
the cutoff interaction.  For consistency one should use the same form
of interaction between all the particles, but it turns out that if we
apply our new interaction to a system of quantum mechanical electrons
and classical ions then it reduces to using the Ewald interaction for
the terms involving the ions while the cutoff interaction applies only
to the electron-electron interactions.  Note that the cutoff
interaction is formulated independently of QMC itself and may be used
with other techniques for periodic systems.\cite{ourfinitesize}

\subsection{The trial wave function}

The choice of trial wave function is of critical importance for VMC
and DMC calculations.  We have used a standard Slater-Jastrow form:

\begin{eqnarray}
\label{eqn:detjaschi}
\Psi_T({\bf r}_1,\ldots,{\bf r}_N) & = 
D^{\uparrow}({\bf r}_1,\ldots,{\bf r}_\frac{N}{2})
D^{\downarrow}({\bf r}_{\frac{N}{2}+1},\ldots,{\bf r}_N) &   \nonumber \\ 
&\times\exp\!\left(\sum_{i=1}^{N}\chi({\bf r}_i) -\sum_{i < j}^{N}u(r_{ij})
          \right)\;, &
\end{eqnarray}

\noindent where the spin-up and spin-down Slater determinants,
$D^{\uparrow}$ and $D^{\downarrow}$, are multiplied by a Jastrow
factor which contains a one-body $\chi$ function and two-body
correlation factor, $u$.  Our $\chi$-function has the full symmetry of
the diamond structure and is expressed as a Fourier series containing
6 inequivalent, non-zero, parameters.  We used spherically symmetric
parallel and antiparallel spin $u$-functions,\cite{ouroptim} which
satisfy the electron-electron cusp conditions\cite{tk:cusp} and
contain a total of 16 parameters.  The optimized parameter values were
obtained by minimizing the variance of the energy.\cite{ouroptim}

The spin-up and spin-down Slater determinants were formed from
single-particle orbitals obtained from an LDA calculation employing
the same pseudopotential as in the QMC calculations.  The LDA orbitals
were calculated at the $\Gamma$-point of the simulation cell Brillouin
zone using a plane wave basis set with an energy cutoff of 15 Ry.
Although the $\Gamma$-point scheme does not give optimal Brillouin
zone sampling,\cite{ourkpoint} it does preserve the full symmetry of
the system and allows comparison with a wider number of established
results. The $\Gamma$-point of the simulation cell Brillouin zone
unfolds to four inequivalent {\bf k}-points in the primitive Brillouin
zone.  These are: (0,0,0) (the $\Gamma$-point),
(0,0,$\frac{2}{3}$)$\frac{2 \pi}{a}$ (a point along the $\Delta$ axis,
hereafter referred to as the $\Delta$-point),
(0,$\frac{2}{3}$,$\frac{2}{3}$)$\frac{2 \pi}{a}$ (a point along the
$\Sigma$ axis, hereafter referred to as the $\Sigma$-point), and
($\frac{1}{3}$,$\frac{1}{3}$,$\frac{1}{3}$)$\frac{2 \pi}{a}$ (a point
along the $\Lambda$ axis, hereafter referred to as the
$\Lambda$-point).

It is highly desirable to improve the quality of the trial wave
functions used in QMC calculations.  Improvements to trial wave
functions can be classified into three types: (i) improvement of the
Jastrow factor, (ii) using a linear combination of determinants, and
(iii) improvements in the orbitals forming the determinants. In this
paper we will investigate a possible improvement of type (iii), namely
the use of natural orbitals. The question of which single-particle
orbitals lead to the best approximation to the exact many-body wave
function is still open. Furthermore, this choice fixes the nodal
surface of the trial wave function and therefore determines the
accuracy of the fixed-node approximation. LDA and HF orbitals have
been used successfully in a number of atomic,\cite{ouroptim,cu:varmin}
molecular\cite{cwg:hydrides,al:hydrides} and solid\cite{he:gaas} QMC
calculations, but so far it has not proved possible to perform a
direct optimization of the single-particle orbitals of an extended
system.  A study of first row atoms and
molecules\cite{cu:correl,cf:mcf} showed that lower energies can be
obtained in both VMC and DMC using a trial wave function containing
several determinants obtained from a multi-configuration
self-consistent field (MCSCF) calculation. However, a similar study
for small silicon clusters found that trial wave functions containing
a single determinant of natural orbitals computed within an MCSCF
scheme gave better DMC results than some multi-determinant wave
functions.\cite{jcg:sicluster} This result strongly suggests that the
natural orbitals result in improved nodal surfaces, and motivates our
calculation of the natural orbitals for bulk silicon.

An expansion of a wave function in Slater determinants of natural
orbitals requires a smaller number of terms for a given accuracy than
expansions using other orbitals.\cite{szabo,mcweeny} Calculation of
the natural orbitals is, however, costly, and less expensive schemes
such as natural pair orbitals\cite{pl:natpairorbs,wm:pno1} have been
proposed to improve convergence in quantum chemical calculations.  It
is not clear that orbitals arising in schemes designed to accelerate
convergence of configuration interaction (CI) calculations should give
smaller fixed-node errors in DMC calculations than LDA or HF orbitals.
However, as mentioned above, there is some evidence to suggest that
natural orbitals have this property.  Natural orbitals have not
frequently been computed within fermion QMC, although VMC and DMC
calculations of natural orbitals have been reported for the ground
states of the Li, C, and Ne atoms.\cite{pha:pseudo} No calculations of
natural orbitals for realistic extended fermion systems have appeared
in the literature to date, although for homogeneous systems the
translational symmetry requires the natural orbitals to be plane
waves.

Systematic studies of multi-determinant wave functions in QMC are
lacking for solids. It seems reasonable to assume that
multi-determinant wave functions will have improved nodes, and
therefore give a better representation of the exact wave function, but
there is little direct evidence to support this.
Multi-configurational approaches include correlation effects, but do
so relatively inefficiently - large numbers of terms (configurations)
are usually required to obtain a significant proportion of the
correlation energy.  This form of wave function is unattractive for
QMC as we require an accurate representation of the wave function
which can be rapidly evaluated.  Therefore, we obtain the one-body
density matrix and hence the natural orbitals from a VMC calculation
using a correlated trial wave function, bypassing the need to
determine them using a multi-determinantal calculation.

\subsection{VMC and DMC calculations}
\label{sec:si}

In VMC we compute the expectation value of the Hamiltonian, $\hat{H}$,
or other operator, with a trial wave function, $\Psi_T$.  This method
gives a rigorous upper bound to the exact ground state energy. The
Metropolis algorithm is used to generate electron configurations,
${\bf R}$, distributed according to $|\Psi_T({\bf R})|^2$, and the
energy calculation is performed by averaging the local energy,
$\Psi_T^{-1} \hat{H} \Psi_T$, over this distribution.

In our DMC calculations we use the short-time approximation for the
Green's function with a time step of 0.015 a.u., which has been shown
to give a small time-step error in silicon.\cite{li} Importance
sampling is introduced via the trial wave function, $\Psi_T$.  We make
the fixed node approximation, restricting the nodes of the DMC
solution to be those of the trial wave function.  Approximately
15$\times 10^3$ statistically independent electron configurations were
used and the acceptance/rejection ratio was greater than $99.9 \%$.
The computational cost of this method scales with the third power of
the system size. Exact fermion techniques, such as the release node
QMC and CI methods, have computational requirements increasing
exponentially with the system size and are impractical for the system
sizes used here.

\section{Calculation of the density matrix and natural orbitals}
\label{sec:dm}

\subsection{Density matrix}

The one-body density matrix\cite{pl:densitymatrices} for a normalized
wave function, $\psi$, is defined as

\begin{eqnarray}
  \label{eqn:densitymatrix}
\rho({\bf r},{\bf r}^\prime)& = & 
   N\int \psi^*( {\bf r},{\bf r}_2,\ldots,{\bf r}_{N}) \nonumber \\
 & \times &\psi( {\bf r}^\prime,{\bf r}_2,\ldots,{\bf r}_{N})
   d{\bf r}_2 \ldots d{\bf r}_N\;. 
\end{eqnarray}

\noindent To facilitate calculation we expand the density matrix
in a basis of orbitals, $\phi_i$, leading to

\begin{equation}
  \label{eqn:densitymatrixbasis} \rho({\bf r},{\bf r}^\prime) =
  \sum_{i,j} \rho_{ij}\phi_i({\bf r})\phi^*_j({\bf r}^\prime)\,.
\end{equation}

\noindent We refer to the diagonal elements, $\rho_{ii}$, as the
orbital occupation numbers.  For wave functions consisting of a single
determinant, such as HF or LDA wave functions, the density matrix is
idempotent $(\rho=\rho^2)$ and takes the form of a sum over the
occupied orbitals, i.e.,

\begin{equation}
  \rho({\bf r},{\bf r}^\prime) = 2
  \sum_{i=1}^{N/2} \phi_i({\bf r})\phi^*_i({\bf r}^\prime)\,,
\end{equation}

\noindent so that the occupation numbers are 2 (including spin
degeneracy) for occupied orbitals and 0 for unoccupied orbitals.

We write the matrix elements of the interacting density matrix,
$\rho_{ij}$, as expectation values over the distribution $\left| \psi
\right| ^2$:

\begin{eqnarray}
  \label{eqn:vmcdensitymatrix} \rho_{ij}=N\int \phi^*_i({\bf
  r}_1)\phi_j({\bf r}^\prime) \frac{\psi( {\bf r}^\prime,{\bf
  r}_2,\ldots,{\bf r}_{N})}{\psi( {\bf r}_1,\ldots,{\bf r}_{N})} & &
   \nonumber \\
  \times \left| \psi( {\bf r}_1,\ldots,{\bf r}_{N})\right| ^2 d{\bf r}^\prime
  d{\bf r}_1\ldots d{\bf r}_N\,. & &
\end{eqnarray}

\noindent The permutation symmetry allows us to rewrite this in a way
which is efficient for Monte Carlo evaluation. Denoting the average
over the distribution, $\left| \psi \right| ^2$, as
$\left<\ldots\right>_{|\psi|^2}$, the Monte Carlo expectation value is
written as

\begin{eqnarray}
  \label{eqn:vmcdm}
\rho_{ij}= \left<\,\sum_{n=1}^N \int  \phi^*_i({\bf r}_n)\phi_j({\bf
  r}^\prime) \frac{\psi(\ldots,{\bf r}^\prime,\ldots)}{\psi(
  \ldots,{\bf r}_n,\ldots)} d{\bf r}^\prime \right>_{|\psi|^2},
\end{eqnarray}

\noindent so that $N$ values are accumulated at each step along the
VMC walk. The integral over $d{\bf r}^\prime$ is performed by summing
over a grid of uniform spacing whose origin is chosen randomly for
each electron configuration. The same grid in ${\bf r}^\prime$ is used
for each term in Eq.~\ref{eqn:vmcdm}, which further reduces the
computational cost. We tested a series of grid sizes for the ${\bf
r}^\prime$ integral, using identical configurations for each grid size
to obtain a correlated sampling estimate of the difference between the
integrals. We found that a grid containing 125 points in the
simulation cell sampled the integral with sufficient accuracy.

Provided that the density of points in the ${\bf r}^\prime$ integral
is kept constant, the statistical error in the individual elements of
the density matrix for a given number of statistically independent
configurations is approximately independent of system size.

We used a basis set consisting of the lowest energy LDA orbitals at
the 27 {\bf k}-points in the primitive Brillouin zone. We tested the effect
of varying the number of orbitals in the basis. We found that
approximately 40 orbitals per {\bf k}-point were sufficient, although to
retain the symmetry we included all members of a degeneracy, so that
the actual number used was either 39 or 40 orbitals, depending on the
{\bf k}-point.  The normalization used in Eq.~(\ref{eqn:densitymatrix})
requires that

\begin{equation}
 N ={\rm T\!r}\rho \; ,
\end{equation}

\noindent which provides a practical test for the completeness of the
basis set. The total occupation of the matrix (Tr$\rho$) was 215.9(2),
which is within the statistical error of the number of electrons in
the system, indicating completeness of the basis at the level of the
statistical accuracy obtained. 

For a given number of Monte Carlo moves, the best statistics are
obtained by accumulating all non-zero matrix elements and applying the
symmetry afterwards.  However it is computationally very expensive to
accumulate all of them, and we found that a more efficient procedure
was to accumulate only the independent non-zero matrix elements. The
basis set of LDA orbitals are basis functions of the unitary
irreducible representations of the symmetry group $O_h^7$. Using the
``orthogonality condition for matrix representations''\cite{corn:gtp1}
we inferred that elements involving products of orbitals from
inequivalent {\bf k}-points and of differing representations are
zero. We ensured that every occurrence of a given representation was
identical, so that products between functions belonging to different
rows were orthogonal. This procedure reduced the total number of
independent and non-zero matrix elements, $\rho_{ij}$, from 42094 to
582 elements.

These matrix elements were sampled using approximately 6.6$\times
10^5$ statistically independent configurations. The correlation
lengths along the VMC walks of both the local energy and density
matrix were found to be essentially the same.

\subsection{Results for the density matrix}

\begin{figure}
\begin{center}
\epsfxsize=8cm \epsfbox{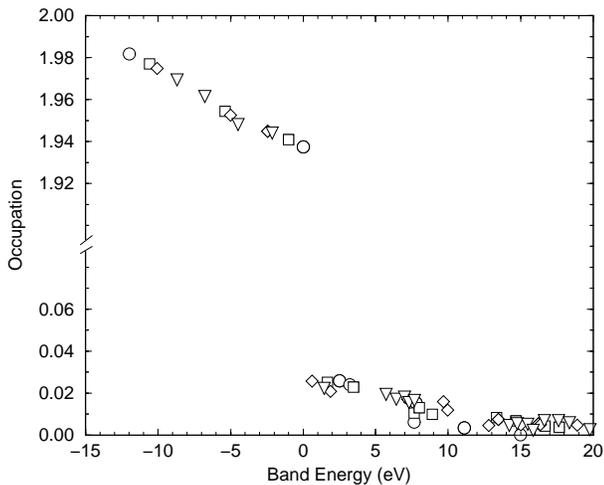}
\end{center}
\caption{Occupation numbers (diagonal elements of the density
  matrix in the basis of LDA orbitals) plotted against the LDA energy
  for each {\bf k}-point. These are the $\Gamma$-point($\circ$),
  $\Delta$ ($\Diamond$), $\Sigma$ ($\bigtriangledown$), and $\Lambda$
  ($\Box$), points on the $\Delta$, $\Sigma$, and $\Lambda$ axes
  respectively.  The statistical error bars are approximately equal to
  the sizes of the symbols for the conduction band states and are
  about 5 times smaller than the symbols for the valence band states.}
\label{fig:occup_vs_ldaen}
\end{figure}

We found the matrix $\rho_{ij}$ to be very nearly diagonal, with
little coupling between LDA orbitals. Double occupancy (spin-up and
down) of orbitals is denoted by the value $2.0$. The maximum
difference between the interacting occupation number, $\rho_{ii}$, and
the LDA occupation number was $0.0625(5)$, which occurred at the
$\Gamma_{25'}$ state at the top of the valence band. The magnitude of
the largest off-diagonal matrix element was $0.014(1)$, which is of
similar order to the occupation number of the lowest unoccupied
orbitals. The fractional errors in occupation numbers for orbitals of
low occupation were large in comparison to those of high occupation.
We found that $97.6\%$ of the total occupation of the density matrix
is contained within the four occupied LDA bands at each {\bf k}-point,
and $99.0\%$ is obtained within the first ten bands.  In
Fig.~\ref{fig:occup_vs_ldaen} we plot the occupation numbers against
the LDA band energies. The occupation numbers decrease almost linearly
with increasing LDA energy for both the occupied and unoccupied bands.

\begin{figure}
\begin{center}
    \epsfxsize=8cm \epsfbox{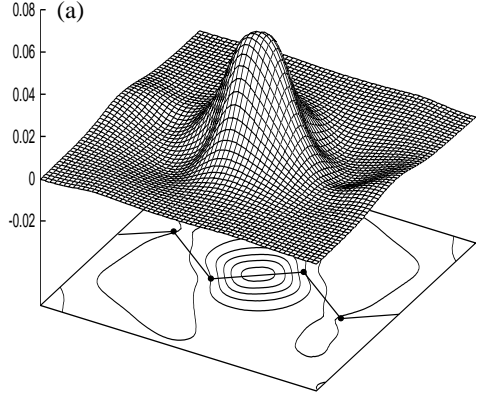}
    \epsfxsize=8cm \epsfbox{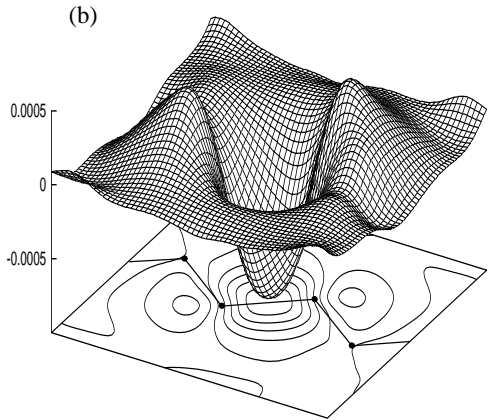}

\end{center}
\caption{(a) The VMC one-body density matrix, $\rho_{\rm VMC}({\bf
r},{\bf r}^\prime)$ and (b) $\rho_{\rm VMC}({\bf r},{\bf
r}^\prime)-\rho_{\rm LDA}({\bf r},{\bf r}^\prime)$, in the (110) plane
passing through the atoms with ${\bf r}$ fixed at the bond
center. $\rho$ is normalized such that $\rho({\bf r},{\bf r})=n({\bf
r})$, the charge density at the bond center. The silicon atoms and
bonds are shown schematically.}
\label{fig:denmat_realspace}
\end{figure}

In Fig.~\ref{fig:denmat_realspace} we show the density matrix,
$\rho({\bf r},{\bf r}^\prime)$, in the (110) plane, and the
differences between the VMC and LDA matrices. The coordinate ${\bf r}$
is fixed at the center of a covalent bond, and ${\bf r}^\prime$ ranges
over the (110) plane passing through the atomic positions. The density
matrix consists of an asymmetric central peak, reduced in width along
the bonding direction. A longer ranged structure is present in areas
of high valence charge density, smaller by approximately one order of
magnitude than the peak.

The VMC value for the peak in the density matrix on the bond center at
${\bf r} \! = \!  {\bf r}^\prime$ is $1.7\%$ smaller than the LDA
value. The VMC density matrix has a larger magnitude around the
neighboring silicon ions than the LDA density matrix, which
consequently has a slightly smaller range.  We also examined the
density matrix in interstitial regions, where we found more structure
to be present.  Again, the LDA and VMC results were very similar, with
small differences between the two cases arising principally from the
differing charge densities.

To investigate the effect of using a finite size simulation cell we
compared the LDA density matrix computed for 3$\times$3$\times$3 and
4$\times$4$\times$4 {\bf k}-point meshes, corresponding to simulation
cells containing 54 and 128 atoms respectively. We found that the
central peak was largely unchanged and the longer ranged structure was
in qualitative agreement. The central maximum in the density matrix in
Fig.~\ref{fig:denmat_realspace} is at the point ${\bf r} \!  = \! {\bf
r}^\prime$ and its magnitude is directly proportional to the valence
charge density at that point, which differed by $4.9\%$ between the
two simulation cell sizes.  We expect the finite size effects in the
QMC calculations broadly to follow those in the LDA, as we have found
for the total energies.\cite{ourfinitesize}

In exact Kohn-Sham density functional theory the total energy can be
written entirely in terms of the one-body density matrix of the
Kohn-Sham orbitals, whereas in a fully interacting system both the
one-body density matrix and the pair-correlation function are
required.  Results for the pair-correlation function from accurate
correlated wave functions and LDA calculations are extremely
different.\cite{ourpairprl} Exact Kohn-Sham density functional theory
reproduces the exact charge density and therefore exactly reproduces
the diagonal ${\bf r}={\bf r}^{\prime}$ part of the density matrix.
The off-diagonal part of the exact Kohn-Sham and interacting density
matrices are not required to be the same.  In silicon we expect the
LDA to give a good approximation to the exact Kohn-Sham density
matrix.  For this system our results show that the {\it entire}
density matrices are very similar in VMC and LDA.

\subsection{Natural Orbitals}

The natural orbitals were obtained by diagonalizing the density matrix
in the basis of the LDA orbitals.  An assessment of the statistical
errors in the eigenvalues and eigenvectors was made by subjecting the
matrix to random perturbations of order the statistical error.  The
eigenvalues varied by up to $\pm 0.0004$ on application of the small
perturbations.  All the calculated eigenvalues, $\lambda_i$, of the
density matrix lie in the range $0 \! \le \!  \lambda_i \! \le \! 2$,
as is required.\cite{pl:densitymatrices} Identical results were
obtained when elements within statistical error of zero were
explicitly zeroed. The overlap of the space occupied by the LDA
orbitals and the corresponding natural orbitals is measured by the
absolute value of the determinant of the matrix of overlaps between
these two sets of vectors.  This gave a value of 0.9948, indicating
that the spaces spanned are almost the same.

The eigenvalues of the density matrix for the ``occupied'' natural
orbitals were very slightly larger than the corresponding matrix
elements $\rho_{ii}$ (by about $0.001$). Consequently, the
eigenvalues of the ``unoccupied'' natural orbitals were very slightly
decreased, so that ${\rm T\!r} \rho$ is invariant. Therefore a plot of
the eigenvalues of the density matrix would be indistinguishable from
Fig.~\ref{fig:occup_vs_ldaen}, which shows the diagonal elements of
the density matrix in the basis of LDA orbitals.

\subsection{DMC Calculations}

As well as the LDA and VMC calculations, we performed fixed node DMC
calculations with trial wave functions of the form of
Eq.~(\ref{eqn:detjaschi}), using LDA and natural orbitals to form the
determinants. Re-optimization of the Jastrow and $\chi$ functions to
improve sampling efficiency in the DMC calculation was found to be
unnecessary. The resulting energies were $-107.59$ eV (LDA),
$-107.69(1)$ eV (VMC with LDA orbitals), $-107.71(1)$ eV (VMC with
natural orbitals), $-108.10(1)$ eV (DMC with LDA orbitals), and
$-108.09(1)$ eV (DMC with natural orbitals).  The VMC wave function
appears to show a very slight improvement with natural orbitals
compared with LDA orbitals. However, to within statistical accuracy,
the DMC energies obtained with LDA and natural orbitals are the same.
This indicates that the nodal surfaces given by the LDA and natural
orbitals are of the same quality.

\subsection{DMC Comparison of LDA and HF Orbitals}

In light of these results it is interesting to compare the quality of
the nodal surfaces obtained with LDA and HF orbitals, which are both
commonly used in the determinantal parts of trial wave functions for
QMC calculations. 

We investigated this by performing DMC calculations in silicon with an
fcc simulation cell containing 16 atoms. The smaller simulation cell
enabled a large number of independent configurations to be obtained
rapidly.  Wave functions expanded in a basis of atom-centered
Gaussians were obtained from the HF and DFT code\cite{crystal95manual}
CRYSTAL95.  We took special care to ensure that the LDA and HF
calculations were done in equivalent ways to try and eliminate any
bias in the comparison.  A basis set of four uncontracted $sp$
functions and one $d$ polarization function per pseudo-atom was
optimized separately for each calculation.  The quality of the basis
set is high - to obtain the same energy within a plane wave
calculation would require a basis set cutoff of 12.5 Ry.  We used the
same non-local LDA pseudopotential as in our other calculations.  In
both calculations we used the same $u$ and $\chi$ functions and
performed DMC simulations with an average population of 640 walkers,
performing approximately $6.7\times 10^5$ walker moves.  We obtained
DMC total energies of -107.488(3) eV per atom and -107.464(3) eV per
atom for the LDA and HF guiding wave functions respectively, using the
Ewald interaction in the many-body Hamiltonian.

The walker energies were approximately normally distributed. Using a
conventional t-test, the 95\% confidence interval on the difference in
energies obtained was $0.002 - 0.046$ eV per atom, showing that for
this system it is very likely that the DMC energy from a determinant
of LDA orbitals is lower than that from a determinant of HF orbitals.
Therefore, for this system, a determinant of LDA orbitals has a
marginally better nodal structure than a determinant of HF orbitals.

\section{Excitation Energies}
\label{sec:ee}

\subsection{Excited State Calculations}

The calculation of excited state energies in solids using QMC methods
is a fairly new area of research.  Significant successes have been
achieved using direct methods, in which separate QMC calculations are
performed for the ground and excited states, and the excitation energy
is calculated as the energy
difference.\cite{lm:nitrogen,lm:compcarb,ourbandgapletter} In these
direct methods a QMC calculation must be performed for each
excitation.  In contrast, for the method described here a large number
of excitation energies are obtained from a single QMC calculation
involving averages over the ground state wave function.

\subsection{The Extended Koopmans' Theorem}

Our method for determining excitation energies corresponds to a QMC
formulation of the Extended Koopmans' Theorem (EKT) derived
independently by Smith, Day and Garrod\cite{owd:ektpotl} and by
Morrell, Parr and Levy.\cite{mmm:ekt} The EKT is closely related to
the earlier work of Feynman\cite{feynman} on calculating excitation
energies in the superfluid state of He$^4$, although the quantum
chemists appear to have developed the theory independently. The EKT
has been shown to give very good excitation energies for simple
molecular systems,\cite{rcm:ektexactness} and has been applied to
atomic and diatomic systems.\cite{owd:ekt2,la:ektlih} It appears
particularly well suited to QMC calculations in which explicitly
correlated many-body wave functions are used. Here we review the
derivation following Ref.~\onlinecite{owd:ektpotl} and present our QMC
formulation.

\subsubsection{Valence Band Energies}

In this method the band energies are calculated as ionization
energies.  We start with an approximation to the normalized ground
state wave function, $\psi^N$.  The wave function for the $N$-1
electron system is approximated by the Ansatz of eliminating an
orbital from $\psi^N$:

\begin{equation}
  \label{eqn:psiN-1} 
 \psi^{N-1}({\bf r}_2,\dots,{\bf r}_N)=\int u_v^*({\bf
 r}_1)\psi^{N}({\bf r}_1,\dots,{\bf r}_N)d{\bf r}_1\;.
\end{equation}

\noindent The valence orbital to be eliminated, $u_v$, will be
determined variationally.  This Ansatz is reminiscent of a
quasiparticle wave function for the excited state, although the
formulation is for $N$-1 particle eigenstates of the Hamiltonian.
Expressing Eq.~(\ref{eqn:psiN-1}) in second quantization yields

\begin{equation}
  |\psi^{N-1}>=\hat{\mathcal{O}}_v|\psi^N>\,,
\end{equation}
\noindent where $\hat{\mathcal{O}}_v$ is the destruction operator for
the state $u_v$.

The ionization energy is given by the difference in the expectation
values of the Hamiltonian calculated with the $N$ and $N$-1 electron
wave functions:\cite{mmm:ekt,owd:ekt1}

\begin{equation}
  \label{eqn:ektve1} \epsilon_v=\langle\psi^N| \hat H| \psi^N \rangle-
    {{\langle \hat{\mathcal{O}}_v\psi^N| \hat
    H|\hat{\mathcal{O}}_v\psi^N \rangle}
    \over{\langle\hat{\mathcal{O}}_v\psi^N|
    \hat{\mathcal{O}}_v\psi^N \rangle}} \,.
\end{equation}

\noindent If $\psi^N$ is an eigenfunction of $\hat H$,
Eq.~(\ref{eqn:ektve1}) may be written as

\begin{equation}
  \label{eqn:ektveop} \epsilon_v=-{{ \langle \psi^N|
  \hat{\mathcal{O}}^\dagger_v[\hat
  H,\hat{\mathcal{O}}_v]|\psi^N\rangle } \over{\langle \psi^N|
  \hat{\mathcal{O}}^\dagger_v\hat{\mathcal{O}}_v|\psi^N\rangle }}\,.
\end{equation}

\noindent The denominator in Eq.~(\ref{eqn:ektveop}) is the one-body
density matrix. We now expand in a set of orbitals, $\{\phi_i\}$, so
that $u_v({\bf r})= \sum c_{iv} \phi_i({\bf r})$, and
$\hat{\mathcal{O}}_v=\sum c_{iv} \hat{a}_i$, where $\hat{a}_i$ is the
destruction operator for $\phi_i$.  The condition for a stationary
value of $\epsilon_v$ generates a secular equation

\begin{equation}
  \label{eqn:occmatrix} ({\bf V}^v-\epsilon_v{\bf S}^v){\bf c}_v={\bf
  0}\,.
\end{equation}

\noindent The matrix ${\bf S}^v$ is the one-body density matrix, and
the elements of ${\bf V}^v$ are $V^v_{ij}=\langle
\psi^N|\hat{a}^\dagger_j[\hat H,\hat{a}_i]|\psi^N\rangle$, where

\begin{eqnarray}
  V^v_{ij}&=&N \int \phi_i({\bf r}_1)\phi^*_j({\bf r}^\prime) \psi^*({\bf
  r}_1,\dots,{\bf r}_N) \nonumber \\
 & &\times \hat{H}_1 \psi({\bf r}^\prime,{\bf
  r}_2,\dots,{\bf r}_N)d{\bf r}^\prime d{\bf r}_1\ldots d{\bf r}_N\,.
\end{eqnarray}

\noindent $\hat{H}_1$ consists of the terms in the $N$-electron
Hamiltonian of Eq.~(\ref{eqn:hamil}) involving coordinate ${\bf r}_1$,
so that

\begin{equation}
  \hat{H}_1= \hat{h}_1+\sum_{j \neq 1}^Nv({\bf r}_1,{\bf r}_j)\,,
\end{equation}

\noindent where $\hat{h}$ consists of the one-body kinetic energy
operator and ionic potential, including both local and non-local
pseudopotential components, and $v$ is the electron-electron
interaction potential.

If a HF wave function is used for $\psi^N$ and the density matrix is
expanded in a basis set of HF orbitals, then $V^v_{ij}$ reduces to a
matrix with the HF $N$-particle eigenvalues on the occupied part of
the diagonal, and zeros everywhere else. The resulting excitation
energies are those given by the well-known Koopmans'
theorem.\cite{koopman} The contents of the EKT method are now
reasonably clear.  The method consists of a quasiparticle-like Ansatz
for the wave function of the $N$-1 particle system, which is used to
calculate the ionization energies of the system.  Electron
correlations are included, but no allowance is made for relaxation of
the other orbitals in the presence of the excitation.  Although this
relaxation can be important in small systems, it is expected to
be much less important for excitations in extended systems such as the
silicon crystal studied here.

\subsubsection{Conduction Band Energies}

An analogous theory exists for the conduction band energies.  The wave
function for the $N$+1 electron system is approximated by the Ansatz
of adding an orbital to $\psi^N$:

\begin{equation}
  \label{eqn:psiN+1}
  \psi^{N+1}({\bf r}_0,\dots,{\bf r}_N)=
       \hat{\mathcal{A}}u_c({\bf r}_0)\psi^N({\bf r}_1,\dots,{\bf r}_N)\,,
\end{equation}

\noindent where $\hat{\mathcal{A}}$ is the antisymmetrizer and the
orbital $u_c$ is to be determined variationally.  In second
quantization we have

\begin{equation}
    |\psi^{N+1}>=\hat{\mathcal{O}}^\dagger_c|\psi^N>\;.
\end{equation}

\noindent The excitation energies $\epsilon_c$ are defined by

\begin{equation}
  \label{eqn:ektceop} \epsilon_c={{ \langle \psi^N|
  \hat{\mathcal{O}}_c[\hat
  H,\hat{\mathcal{O}}^\dagger_c]|\psi^N\rangle} \over{\langle\psi^N|
  \hat{\mathcal{O}}_c\hat{\mathcal{O}}^\dagger_c|\psi^N\rangle}}\,.
\end{equation}

\noindent Expanding in a set of orbitals gives $u_c({\bf r})= \sum
c_{ic} \phi_i({\bf r})$ and $\hat{\mathcal{O}}^\dagger_c=\sum c_{ic}
\hat{a}^\dagger_i$. The coefficients $c_{ic}$ are the solutions of the
secular equation

\begin{equation}
  \label{eqn:unoccmatrix} ({\bf V}^c-\epsilon_c{\bf S}^c){\bf
  c}_c={\bf 0}\,,
\end{equation}

\noindent where the matrix elements are $V^c_{ij}=\langle
\psi^N|\hat{a}_i[\hat H,\hat{a}^\dagger_j]|\psi^N\rangle$ and
$S^c_{ij}=\langle \psi^N|\hat{a}_i\hat{a}^\dagger_j|\psi^N \rangle =
\delta_{ij}-\rho_{ij}$.

It is instructive to introduce a new potential with matrix elements
$V_{ij}=V^v_{ij}+V^c_{ij}$:

\begin{eqnarray}
\label{eqn:vij}
V_{ij}&=\int& \phi_i({\bf r}_0)\hat{h}_0 \phi^*_j({\bf r}_0)d{\bf r}_0
\nonumber \\ &+N\int& \phi_i({\bf r}_0)\phi^*_j({\bf r}_0)v({\bf
r}_0,{\bf r}_1)\nonumber \\ & & \times |\psi({\bf r}_1,\dots,{\bf
r}_N)|^2d{\bf r}_0 d{\bf r}_1 \ldots d{\bf r}_N \nonumber \\ &-N\int&
\phi_i({\bf r}_0)\phi^*_j({\bf r}_1)v({\bf r}_0,{\bf r}_1) \psi({\bf
r}_0,{\bf r}_2,\dots,{\bf r}_N) \nonumber \\ & & \times \psi^*({\bf
r}_1,\dots,{\bf r}_N) d{\bf r}_0 d{\bf r}_1 \ldots d{\bf r}_N \,.
\end{eqnarray}

\noindent If we use a HF wave function and expand in a basis of HF
orbitals, $V_{ij}$ reduces to a matrix with the HF $N$-particle
eigenvalues on the diagonal, and zeros everywhere else. In this case
one can readily identify the second and third terms in
Eq.~(\ref{eqn:vij}) as, respectively, the Hartree and exchange terms.
(Similarly $V^c_{ij}$ reduces to the HF energy eigenvalues on the
unoccupied part of the diagonal.)

\subsubsection{VMC formulation of the EKT}

We accumulate the matrix elements of $V^v_{ij}$ and $V_{ij}$,
subsequently forming the matrix $V^c_{ij} = V_{ij} - V^v_{ij}$.  The
matrix elements, $V^v_{ij}$, are given by 

\begin{eqnarray}
\label{eqn:occelem}
  \lefteqn{V^v_{ij}=N\int \phi_i({\bf r}_1)\phi^*_j({\bf r}^\prime)
  \frac{\hat{H}_1 \psi({\bf r}_1,\dots,{\bf r}_N)} {\psi({\bf
  r}_1,\dots,{\bf r}_N)} } \nonumber \\
  & & \times \frac{\psi( {\bf r}^\prime,{\bf
  r}_2,\ldots,{\bf r}_{N})} {\psi( {\bf r}_1,\ldots,{\bf r}_{N})}
 |\psi({\bf r}_1,\dots,{\bf r}_N)|^2 d{\bf r}^\prime d{\bf r}_1\ldots
  d{\bf r}_N\,.
\end{eqnarray}

\noindent As before, we use the permutation symmetry to write this as

\begin{eqnarray}
  \label{eqn:vmcvij}
V^v_{ij}=
  \left<\,\sum_{n=1}^N \int  \phi^*_i({\bf r}_n)\phi_j({\bf r}^\prime)
\frac{\hat{H}_n \psi({\bf r}_1,\dots,{\bf r}_N)}{\psi({\bf
  r}_1,\dots,{\bf r}_N)}\right. & &  \nonumber \\
\left. \times \frac{\psi(\ldots,{\bf r}^\prime, \ldots)}{\psi( \ldots,{\bf r}_n,\ldots)} 
d{\bf r}^\prime \right>_{|\psi|^2}, & &
\end{eqnarray}

\noindent so that $N$ values are accumulated at each step. The terms
contributing to $\hat{H}_n\psi/\psi$ are already available in a VMC
calculation, allowing $V^v_{ij}$ to be accumulated with virtually no
additional cost beyond that required for the density matrix.  An
analogous VMC formulation for calculating $V_{ij}$ was used.  The
single particle terms, $\hat{h}_0$, appearing in the first term in
Eq.~(\ref{eqn:vij}) are evaluated directly without using Monte Carlo
integration.  We found that using Monte Carlo integration for all the
terms resulted in a small increase in the variance of the matrix
elements. Therefore we prefer to calculate the $\hat{h}_0$ terms
directly. The matrix elements $V^v_{ij}$ and $V_{ij}$ were accumulated
at the same time as the elements $\rho_{ij}$.  The full crystal
symmetry was again used, which reduces the number of matrix elements
which must be accumulated and ensures the correct symmetry in the
presence of statistical noise.

\subsection{Results for excitation energies}

\subsubsection{Full EKT results}

The correlation lengths along the VMC walks for $V^v_{ij}$ and
$V_{ij}$ were found to be similar to those of the local energy and
density matrix, and the distribution of statistical errors was similar
to that for the density matrix. The elements of $V^v_{ij}$ and
$V^c_{ij}$ with the largest statistical errors were the diagonal
elements. The statistical error bars on these elements are estimated
to be $\pm 0.2$ eV.

The matrix Eqs.~\ref{eqn:occmatrix} and \ref{eqn:unoccmatrix} were
diagonalized using a generalized eigenvalue solver. The ratios of the
statistical error bars to the mean values were significantly larger
for the diagonal elements of $V^v_{ij}$ and $V^c_{ij}$ than for
$\rho_{ij}$. Just as for the density matrix, small perturbations were
applied to estimate the statistical errors in the eigenvalues and
eigenvectors. The numerical stability of the diagonalization was
improved by explicitly zeroing elements of $V^v_{ij}$ and $V^c_{ij}$
within statistical error of zero.  The accuracy of the eigenvalues was
further verified by gradually increasing the number of bands in the
diagonalization procedure.  The valence and low lying conduction band
energies were stable to within $\pm 0.4$ eV.

\begin{figure}
  \begin{center}
    \epsfxsize=8cm \epsfbox{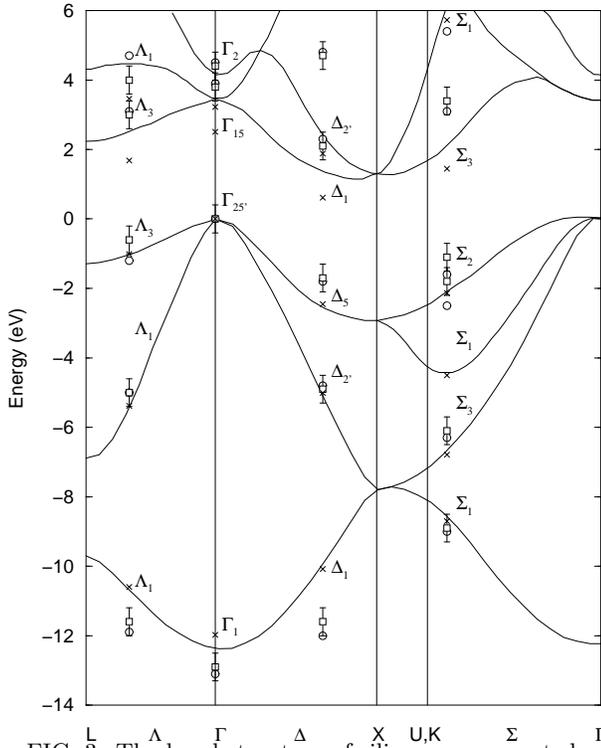}
    \caption{The band structure of silicon, as computed within the VMC
      Extended Koopmans' Theorem using the cutoff interaction
      ($\Box$) and the Ewald interaction ($\circ$), and within the
      LDA ($\times$). The results for the cutoff interaction are shown with
      statistical error bars. The statistical error bars for the Ewald
      results have been omitted for clarity, but they are the same
      size as for the cutoff interaction. As a guide to the eye, the
      empirical pseudopotential data of
      [\protect\onlinecite{jrc:siemppseudo}] is shown (solid lines),
      which is in good agreement with the available experimental
      data.}
\label{fig:bandstructure} 
\end{center}
\end{figure}

The resulting band energies are given in Table \ref{tab:ektvals} and
in Fig.~\ref{fig:bandstructure}, with the energy at the top of the
valence band set to zero.  Of the {\bf k}-points computed here, the
available experimental data (Exp) is limited to the $\Gamma$-point.
Because of this we also give empirical pseudopotential (Emp)
data\cite{jrc:siemppseudo} in Table \ref{tab:ektvals} and in
Fig.~\ref{fig:bandstructure}, which should provide a good
interpolation between this data.  Results for the Ewald and cutoff
interactions are in good agreement, indicating that the finite size
errors are not significant at the level of statistical accuracy
obtained here.  The energies are in good qualitative agreement with
the empirical data at all {\bf k}-points except for the upper
$\Sigma_1$ valence band state, and the $\Delta_{2^\prime}$ conduction
band state.  The source of the error is principally the value of the
diagonal matrix elements for these states, $V^v_{ii}$ and $V^c_{ii}$
(see next section).

The EKT band energies at the $\Gamma$-point are in good agreement with
the available experimental data, and are also in good agreement with
the DMC data from direct calculations of the excitation
energies.\cite{ourbandgapletter} The EKT valence band width of
$12.9(6)$ eV is smaller than the value of $13.6(3)$ eV obtained from
the DMC calculations and is in good agreement with the experimental
value of $12.5(6)$ eV. In comparison the HF data shows the well known
overestimation of band gaps and band widths which is due to the
neglect of correlation energy, the LDA gives excellent valence band
energies while the conduction bands are too low in energy by 0.7-1.0
eV, and the $GW$ data is in very good agreement with experiment.

The EKT is a formulation for the eigenstates of the $N$-1 and $N$+1
electron systems, while the direct method is aimed at calculating the
eigenstates of the $N$ electron system.  The EKT and direct results
should therefore differ by the exciton binding energy, but this energy
is small for silicon and cannot be resolved at the level of
statistical accuracy obtained. Depending on the application one would
like to be able to choose whether to include excitonic effects, so
that it is advantageous to have the different QMC techniques
available. Clearly further refinement of excited state QMC methods is
required, but the results from the EKT and direct approaches are
promising for the study of more strongly correlated systems, for which
the LDA and $GW$ methods are less reliable.

\subsubsection{Diagonal approximations}

If we neglect the off-diagonal element of $V^v_{ij}$, $V^c_{ij}$, and
$\rho_{ij}$ then the valence and conduction band energies can be
approximated by

\begin{equation}
  \label{eqn:smapp} 
  \epsilon_{iv}^{\rm {DEKT}} = \frac{V^v_{ii}}{\rho_{ii}}\;,\;
  \epsilon_{ic}^{\rm {DEKT}} = \frac{V^c_{ii}}{1-\rho_{ii}} \,,
\end{equation}

\noindent where the superscript DEKT denotes the diagonal
approximation to the EKT. This approximation has been used within VMC
by Fahy et al.\cite{sf:pairprl} to calculate quasihole energies in
silicon and by Tanaka\cite{st:nio2} to calculate quasihole energies in
NiO. This approximation has the computational advantage that far fewer
matrix elements are required, and also that the problems of
statistical noise are reduced, because the values of only two matrix
elements enter the calculation of each band energy. However the
results are basis set dependent, and could differ significantly
between, for example, a HF and LDA basis.

If we use a HF wave function and expand in a basis set of HF orbitals
then the matrices $V^v_{ij}$, $V^c_{ij}$, and $\rho_{ij}$ are all
diagonal, and consequently the full EKT and the DEKT are equivalent.
In general, for correlated wave functions, the DEKT gives neither an
upper nor lower bound to the energy obtained from the full EKT.
Comparison with the data from the DEKT is also made in
Table~\ref{tab:ektvals}. The DEKT values are close to the full EKT
values, including the two cases mentioned above where the agreement
with experiment is poor.  The DEKT works quite well for the valence
bands and slightly less well for the conduction bands, because the
off-diagonal matrix elements of $V^c_{ij}$ coupling the unoccupied
states are more significant.  In similar VMC calculations for silicon
using the DEKT and a simulation cell containing 64 electrons, Fahy et
al.\cite{sf:pairprl} calculated a valence band width of 14.5(4) eV,
which is larger than both our value of 13.3(2) eV and the experimental
value of 12.5(6) eV.  Fahy et al.\cite{sf:pairprl} also obtained
occupation numbers of 1.96(4) and 1.92(3) for the $\Gamma_1$ and
$\Gamma_{25^\prime}$ valence band states, respectively, which are
within error bars of our results of 1.9817(2) and 1.9375(5).

The scaling with system size of the diagonal approximation is very
advantageous compared with direct evaluation of excitation energies.
In direct methods, the fractional energy change due to the promotion
of an electron is considered.  This energy change is inversely
proportional to the number of electrons in the system, i.e., a
`$\frac{1}{N}$' effect.  The precision of the calculation must
therefore be sufficient to resolve the energy change amid the
statistical noise. This requirement is very challenging for small band
gap materials, as the system must also be sufficiently large to
approximate the infinite solid. In the DEKT, the band energies are
computed directly as averages over the square of the ground state wave
function, rather than as the difference between averages over the
squares of the ground and excited state wave functions.  This greatly
improves the sampling statistics.  We have found that the errors in
$V^v_{ii}$, $V^c_{ii}$ and $\rho_{ii}$ scale as the inverse of the
square root of the number of independent electron configurations,
independently of the system size.  For sufficiently large systems, it
therefore requires fewer statistically independent samples in the DEKT
to obtain excitation energies to a given statistical accuracy than
with direct methods.  The scaling behavior of the full EKT is more
difficult to analyze because we are required to diagonalize noisy
matrix equations, where the number of off-diagonal matrix elements
increases as the square of the system size, and where there are
certainly statistical correlations between matrix elements.

\section{Conclusions}

We have calculated the one-body density matrix for the valence
electrons of bulk silicon using QMC techniques. 

In real space, the VMC and LDA density matrices are very similar, the
greatest differences being due to the differing charge densities in
each case.  The natural orbitals, obtained by diagonalizing the
density matrix, very closely resemble the LDA orbitals. The occupation
numbers of the natural orbitals differ significantly from the
non-interacting values, reducing linearly with increasing energy for
the LDA occupied bands.  The occupations are about 3\% lower than the
non-interacting value near the top of the valence band, and above the
Fermi level, the occupation numbers fall slowly to zero.

A DMC calculation for the ground state energy of silicon using a trial
wave function containing a determinant of natural orbitals gives an
energy which is almost identical to that obtained using a determinant
of LDA orbitals. This shows that the quality of the nodal surfaces is
almost identical in each case.  We used DMC calculations to compare
the quality of the nodal surfaces obtained with LDA and HF orbitals,
finding that the LDA orbitals gave a slightly lower DMC energy,
indicating that for the system studied the nodal surface of a
determinant of LDA orbitals is slightly better. We note that a
previous DMC study of small silicon clusters\cite{jcg:sicluster} found
that natural orbitals obtained from a multi-configurational
Hartree-Fock calculation gave a better nodal surface than a single
determinant of HF orbitals.

We have calculated excitation energies in silicon using an extension
of Koopmans' theorem applicable to correlated wave functions. The
Monte Carlo formulation is very similar to that required to obtain the
density matrix. The resulting band energies are in good agreement with
the available experimental data.

The success of the VMC-EKT relies on a cancelation of errors between
the ground and excited state energies.  The wave functions for the
excited states contain the variational freedom of the orbitals $u_v$
and $u_c$.  This variational freedom in the orbitals reduces the
energy of the $N$-1 and $N$+1 electron states and improves the
agreement with experiment.  In the diagonal approximation to the EKT
(DEKT), the $u_v$ and $u_c$ orbitals are fixed and there is no
variational freedom in the excited state wave functions.  We have
found that the DEKT works quite well for silicon using LDA orbitals
for $u_v$ and $u_c$.  The diagonal approximation is exact within HF
theory, so that we expect it to be a good approximation for weakly
correlated systems.

Greater accuracy could be obtained with more accurate trial functions,
or using DMC in the calculation of $V^v_{ij}$, $V^c_{ij}$, and
$\rho_{ij}$. In comparison with direct methods of calculating
excitation energies,\cite{lm:nitrogen,lm:compcarb,ourbandgapletter}
the EKT has the advantage that only a single calculation involving the
ground state is required to obtain many excitation energies.

The EKT involves assumptions about the nature of the excited state
wave functions, but nevertheless is a practical method for calculating
excitation energies including correlation effects for relatively
weakly correlated systems. The diagonal approximation to the EKT has a
very advantageous scaling with system size compared with direct QMC
calculations. This scaling thereby allows the study of excitation
energies in larger systems with a greater efficiency than is possible
with direct techniques.

This work was supported by the Engineering and Physical Sciences
Research Council (UK).  Our calculations are performed on the CRAY-T3D
at the Edinburgh Parallel Computing Centre, and the Hitachi SR2201
located at the University of Cambridge High Performance Computing
Facility.

\narrowtext

\clearpage

\clearpage
\narrowtext

\clearpage
\mediumtext

\begin{table}
\begin{tabular}{ldddddddddc}
Band&\multicolumn{4}{c}{VMC}&DMC$^e$&HF$^f$&LDA$^g$&$GW^h$&Emp$^i$&Exp$^j$\\
                  &EKT$^a$&EKT$^b$&DEKT$^c$&DEKT$^d$& & & & & &\\ \tableline
$\Gamma_{2^\prime}$ &  4.4&  4.5&  5.1&  5.2&  4.6&  9.0&  3.22&  3.89&  4.1&4.23,4.1     \\
$\Gamma_{15}$       &  3.8&  3.9&  4.4&  4.4&  3.7&  8.0&  2.51&  3.36&  3.4&3.40,3.05    \\
$\Gamma_{25^\prime}$&  0.0&  0.0&  0.0&  0.0&  0.0&  0.0&  0.00&  0.00&  0.0&0.00         \\
$\Gamma_{1}$        &-12.9&-13.1&-13.3&-13.4&-13.6&-18.9&-11.98&-11.95&-12.36&-12.5$\pm$0.6\\
$\Lambda_3$         &  4.0&  4.7&  5.4&  5.7&     &  9.0&  3.46&      &  4.5&\\
$\Lambda_1$         &  3.0&  3.1&  3.6&  3.6&     &  6.8&  1.68&      &  2.5&\\
$\Lambda_{3}$       & -0.6& -1.2& -1.3& -1.5&     & -1.7& -1.01&      & -0.9&\\
$\Lambda_1$         & -5.0& -5.0& -5.4& -5.4&     & -6.3& -5.38&      & -5.2&\\
$\Lambda_1$         &-11.6&-11.9&-12.3&-12.5&     &-16.4&-10.60&      &-10.6&\\
$\Delta_{2^\prime}$ &  4.7&  4.8&  5.7&  6.0&     &  7.2&  1.88&      &  2.4&\\
$\Delta_{1}$        &  2.1&  2.3&  2.7&  3.0&     &  5.5&  0.61&      &  1.3&\\
$\Delta_5$          & -1.7& -1.8& -2.4& -2.5&     & -3.8& -2.45&      & -2.5&\\
$\Delta_{2^\prime}$ & -4.9& -4.8& -5.5& -5.7&     & -7.8& -5.01&      & -5.0&\\
$\Delta_{1}$        &-11.6&-12.0&-11.7&-12.2&     &-16.0&-10.08&      &-10.2&\\
$\Sigma_1$          &  6.4&  5.4&  6.7&  6.1&     &     &  5.73&      &  6.0&\\
$\Sigma_3$          &  3.4&  3.1&  4.1&  4.0&     &     &  1.45&      &  2.0&\\
$\Sigma_2$          & -1.1& -1.6& -1.2& -2.0&     &     & -2.14&      & -2.1&\\
$\Sigma_1$          & -1.8& -2.5& -1.9& -2.5&     &     & -4.50&      & -4.6&\\
$\Sigma_3$          & -6.1& -6.3& -6.4& -6.8&     &     & -6.79&      & -6.7&\\
$\Sigma_1$          & -8.9& -9.0& -9.5& -9.8&     &     & -8.70&      & -8.7&\\
\end{tabular}
\caption{Band energies of silicon in eV.
  a,b- VMC-EKT energies using the cutoff and Ewald interactions,
  respectively. The statistical error bars are $\pm0.4$ eV.  c,d-
  Diagonal approximation to the VMC-EKT energies using the cutoff and
  Ewald interactions, respectively. The statistical error bars are
  $\pm0.2$ eV.  e- Direct DMC calculations, with statistical error
  bars of $\pm0.2$ eV, from Ref.
  \protect\onlinecite{ourbandgapletter}.  f- Ref.
  \protect\onlinecite{wdl:sihfbandstruc}.  g- This work.  h- Ref.
  \protect\onlinecite{mr:gw}.  i- Ref.
  \protect\onlinecite{jrc:siemppseudo} j- From the compilation given
  in Ref. \protect\onlinecite{mr:gw}. }
\label{tab:ektvals}
\end{table}
\end{document}